\documentclass[conference]{IEEEtran}

\usepackage{cite}
\usepackage{amsmath,amssymb,amsfonts}
\usepackage{graphicx}
\usepackage{multicol}
\usepackage{multirow}
\usepackage{amsmath,amssymb,amsfonts}
\usepackage{float}
\usepackage{textcomp}
\usepackage{csquotes}
\usepackage{lipsum}
\usepackage{easyReview}
\usepackage{tikz}
\usepackage{amsmath}
\usepackage{amsthm}
\usepackage{amssymb}
\usepackage{subcaption}
\usepackage{makecell}
\usepackage{booktabs}
\usepackage[acronym,toc=false,nonumberlist]{glossaries-extra}
\makenoidxglossaries
\GlsXtrEnableLinkCounting{general,acronym}
\setabbreviationstyle[acronym]{long-short}

\newacronym{adas}{ADAS}{Advanced Driver-Assistance Systems}
\newacronym{adl}{ADL}{Architecture Description Language}
\newacronym{alt}{ALT}{\emph{Altimeter}}
\newacronym{apd}{APD}{\emph{Advanced Pedestrians Detection}}
\newacronym{bcet}{BCET}{Best-Case Execution Time}
\newacronym{bddf}{BDDF}{Bounded Dynamic Dataflow}
\newacronym{bdf}{BDF}{Boolean-controlled Dataflow}
\newacronym{bf}{BF}{Blocking Factor}
\newacronym{biro}{BIRO}{Between Iteration Runtime Oriented}
\newacronym{biso}{BISO}{Between Iteration Statically Oriented}
\newacronym{bnf}{BNF}{Backus-Naur Form}
\newacronym{bpdf}{BPDF}{Boolean Parametric Dataflow}
\newacronym{ca}{CA}{\emph{Control Altitude}}
\newacronym{cam}{CAM}{\emph{Camera}}
\newacronym{cddf}{CDDF}{Cyclo Dynamic Dataflow}
\newacronym{cfdf}{CFDF}{Core Functional Dataflow}
\newacronym{cf-psdf}{CF-PSDF}{Core Functional - Parameterized Synchronous Dataflow}
\newacronym{cg}{CG}{Computation Graph}
\newacronym{co}{Co}{Consistency}
\newacronym{cps}{CPS}{Cyber-Physical System}
\newacronym{cpu}{CPU}{Central Processing Unit}
\newacronym{ct}{CT}{\emph{Control Translation}}
\newacronym{cthr}{CT}{Consumption Threshold}
\newacronym{csdf}{CSDF}{Cyclo-Static Dataflow}
\newacronym{csdfa}{CSDF$^a$}{Cyclo-Static Dataflow with auto-concurrency}
\newacronym{cv-sdf}{CV-SDF}{Computer Vision-Synchronous Dataflow}
\newacronym{cy1}{CY1}{\emph{Control Yaw 1}}
\newacronym{cy2}{CY2}{\emph{Control Yaw 2}}
\newacronym{dec}{Dec}{Decidability}
\newacronym{del}{Del}{Delay}
\newacronym{det}{FuncDet}{Functional Determinism}
\newacronym{dfg}{DFG}{Dataflow Graph}
\newacronym{dfmocc}{DF~MoCC}{Dataflow Model of Computation and Communication}
\newacronym{dmd}{DMD}{\emph{Depth Mask Detection}}
\newacronym{dp}{DP}{Dataflow Process}
\newacronym{dpp}{DPP}{Dynamic-Priority Preemptive}
\newacronym{dpn}{DPN}{Dataflow Process Network}
\newacronym{dsp}{DSP}{Digital Signal Processing}
\newacronym{ebs}{EBS}{\emph{Emergency Brake System}}
\newacronym{edf}{EDF}{Earliest Deadline First}
\newacronym{eft}{EFT}{Earliest Finish Time}
\newacronym{eidf}{EIDF}{Enable-Invoke Dataflow}
\newacronym{esadf}{eSADF}{Exponentially timed Scenario-Aware Dataflow}
\newacronym{et}{ET}{Execution Time}
\newacronym{execwin}{ExecWin}{Execution Windows}
\newacronym{fc}{FC}{Flight Computer}
\newacronym{fcImuInt}{FC-IMU-INT}{\emph{FC - IMU Integration}}
\newacronym{fd}{FD}{\emph{Feature Detection}}
\newacronym{fifo}{FIFO}{First-In First-Out}
\newacronym{fm}{FM}{\emph{Feature Match}}
\newacronym{fp}{FP}{\emph{Filtering Procedure}}
\newacronym{fpga}{FPGA}{Field-Programmable Gate Array}
\newacronym{fpp}{FPP}{Fixed-Priority Preemptive}
\newacronym{frdf}{FRDF}{Fractional Rate Dataflow}
\newacronym{fsm}{FSM}{Finite State Machine}
\newacronym{fsm-sadf}{FSM-SADF}{Finite State Machine-based Scenario-Aware Dataflow}
\newacronym{fsm-psadf}{FSM-PSADF}{Finite State Machine-based Parameterized Scenario-Aware Dataflow}
\newacronym{ft}{FT}{\emph{Feature Tracking}}
\newacronym{freq}{Freq}{Frequency}
\newacronym{gcd}{GCD}{Greatest Common Divisor}
\newacronym{gs}{GS}{Global State}
\newacronym{hcfdf}{HCFDF}{Hierarchical Core Functional Dataflow}
\newacronym{hdf}{HDF}{Heterochronous Dataflow}
\newacronym{hi}{Hi}{Hierarchy}
\newacronym{hpdf}{HPDF}{Homogeneous Parameterized Dataflow}
\newacronym{hsdf}{HSDF}{Homogeneous Synchronous Dataflow}
\newacronym{hsdfa}{HSDF$^a$}{Homogeneous Synchronous Dataflow with auto-concurrency}
\newacronym{hvac}{HVAC}{Heating, Ventilation, and Air Conditioning}
\newacronym{ibsdf}{IBSDF}{Interface-Based Synchronous Dataflow}
\newacronym{idf}{IDF}{Integer-controlled Dataflow}
\newacronym{ifd}{IFD}{\emph{Information Display}}
\newacronym{ildf}{ILDF}{Interval-rate Locally-static Dataflow}
\newacronym{imu}{IMU}{Inertial Measurement Unit}
\newacronym{imuCorr}{IMU-CORR}{\emph{IMU Correction}}
\newacronym{inistep}{IniSteP}{Initial and Steady Phases}
\newacronym{inidisit}{IniDisIT}{Initial and Discard of Initial Tokens}
\newacronym{iot}{IoT}{Internet-of-Things}
\newacronym{it}{IT}{Initial Tokens}
\newacronym{kpn}{KPN}{Kahn Process Network}
\newacronym{la}{La}{Latency}
\newacronym{lcm}{LCM}{\emph{Left Camera}}
\newacronym{ld}{LD}{\emph{Label Decider}}
\newacronym{ldr}{LDR}{\emph{Lidar}}
\newacronym{li}{Li}{Liveness}
\newacronym{lf}{LF}{Lingua Franca}
\newacronym{lst}{LST}{Latest Start Time}
\newacronym{ltl}{LTL}{Linear Temporal Logic}
\newacronym{ma}{MA}{Markov Automata}
\newacronym{mdf}{MDF}{Multi-Dimensional FIFO}
\newacronym{mdsdf}{MDSDF}{Multi-Dimensional Synchronous Dataflow}
\newacronym{me}{Me}{Memory}
\newacronym{mem}{MEM}{Multiple Execution Mode}
\newacronym{milp}{MILP}{Mixed Integer Linear Programming}
\newacronym{mm}{MM}{Meta-Model}
\newacronym{mmlp}{MMLP}{Min-Max Linear Programming}
\newacronym{mocc}{MoCC}{Model of Computation and Communication}
\newacronym{mot}{MOT}{\emph{Motors}}
\newacronym{mpsoc}{MPSoC}{Multi-Processor System-on-Chip}
\newacronym{nasa}{NASA}{National Aeronautics and Space Administration}
\newacronym{nc}{NC}{Navigation Computer}
\newacronym{ncImuInt}{NC-IMU-INT}{\emph{NC - IMU Integration}}
\newacronym{nf}{NF}{\emph{Navigation Filter}}
\newacronym{nsf}{NSF}{National Science Foundation}
\newacronym{obd}{OBD}{\emph{Near-field Obstacle Detection}}
\newacronym{odm}{ODM}{\emph{Odometer}}
\newacronym{ooc}{OOC}{Out-of-Order Consumption}
\newacronym{os}{OS}{Operating System}
\newacronym{pa}{Pa}{Parameters}
\newacronym{pci}{PCI}{Production and Consumption Instants}
\newacronym{pcg}{PCG}{Phased Computation Graph}
\newacronym{pcsdf}{PCSDF}{Parameterized Cyclo-Static Dataflow}
\newacronym{pdd}{PDD}{\emph{Pedestrians Detection}}
\newacronym{ph}{Ph}{Phase}
\newacronym{pimm}{PIMM}{Parameterized and Interfaced Meta-Model}
\newacronym{pisdf}{PISDF}{Parameterized and Interfaced Synchronous Dataflow}
\newacronym{pl}{PL}{\emph{Pseudo Landmarks}}
\newacronym{ppsdf}{ppSDF}{Partially Periodic Synchronous Dataflow}
\newacronym{psdf}{PSDF}{Parameterized Synchronous Dataflow}
\newacronym{psm}{PSM}{Parameterized Set of Modes}
\newacronym{psm-cfdf}{PSM-CFDF}{Parameterized Set of Modes - Core Functional Dataflow}
\newacronym{pu}{PU}{Processor Utilization}
\newacronym{qsc}{QSc}{Quasi-Static Schedule}
\newacronym{rai}{RaI}{Rate as Interval}
\newacronym{rcm}{RCM}{\emph{Right Camera}}
\newacronym{rdf}{RDF}{Reconfigurable Dataflow}
\newacronym{rm}{RM}{Rate Monotonic}
\newacronym{rmd}{RMD}{\emph{Road Mask Detection}}
\newacronym{rmdf}{RMDF}{Real-time Mode-aware Dataflow}
\newacronym{rpn}{RPN}{Reactive Process Network}
\newacronym{sad}{SAD}{State-Aware Dataflow}
\newacronym{sadf}{SADF}{Scenario-Aware Dataflow}
\newacronym{stasch}{StaSch}{Statically Schedulable}
\newacronym{sco}{SCo}{Strong Consistency}
\newacronym{sdf}{SDF}{Synchronous Dataflow}
\newacronym{sp}{SP}{\emph{State Propagation}}
\newacronym{spbdf}{SPBDF}{Synchronous PiggyBacked Dataflow}
\newacronym{spc}{SPC}{\emph{Speed Control}}
\newacronym{spdf}{SPDF}{Schedulable Parametric Dataflow}
\newacronym{ssdf}{SSDF}{Scalable Synchronous Dataflow}
\newacronym{swi}{SWi}{Sliding Windows}
\newacronym{tcsdf}{tCSDF}{Timed Cyclo-Static Dataflow}
\newacronym{th}{Th}{Throughput}
\newacronym{tld}{TLD}{\emph{Traffic Lanes Detection}}
\newacronym{tpdf}{TPDF}{Transaction Parameterized Dataflow}
\newacronym{tsd}{TSD}{\emph{Traffic Sign Detection}}
\newacronym{tsdf}{tSDF}{Timed Synchronous Dataflow}
\newacronym{uppaal}{UPPAAL}{UPPAAL}
\newacronym{vlsi}{VLSI}{Very Large Scale Integration}
\newacronym{vpdf}{VPDF}{Variable-rate Phased Dataflow}
\newacronym{vrdf}{VRDF}{Variable Rate Dataflow}
\newacronym{vsdf}{VSDF}{Synchronous Dataflow for VLSI}
\newacronym{way}{WAY}{\emph{Waypoints}}
\newacronym{wcet}{WCET}{Worst-Case Execution Time}
\newacronym{wiro}{WIRO}{Within Iteration Runtime Oriented}
\newacronym{wiso}{WISO}{Within Iteration Statically Oriented}
\newacronym{wsdf}{WSDF}{Windowed Synchronous Dataflow}
\newacronym{xsadf}{xSADF}{Flexible Scenario-Aware Dataflow}

\usepackage{threeparttable}
\usepackage[ruled,linesnumbered,vlined]{algorithm2e}
\usepackage[capitalize,nameinlink]{cleveref}
\usetikzlibrary{shapes}
\usetikzlibrary{positioning}
\usetikzlibrary{patterns}
\usetikzlibrary{arrows.meta}

\SetKwComment{Comment}{$\triangleright$~}{}
\SetKwProg{While}{while}{:}{}
\SetKwProg{For}{for}{:}{}
\SetKw{Continue}{continue}
\SetKwProg{Def}{def}{:}{}
\SetKwBlock{Begin}{Begin}{}

\tikzset{actor/.style={align=center, draw, rounded corners=.20cm}}
\tikzset{setback/.style={fill=gray!50}}
\tikzset{emphasize/.style={line width=2pt,font=\bfseries}}
\tikzset{virtualactor/.style={align=center, draw, ellipse}}
\tikzset{tapeactor/.style={tape,align=center,draw}}
\tikzset{decider/.style={align=center, draw, regular polygon, regular polygon sides=6}}
\tikzset{arc/.style={thin,-{Stealth}}}
\tikzset{arccontrol/.style={thin,-{Stealth},dashed}}

\def\BibTeX{{\rm B\kern-.05em{\sc i\kern-.025em b}\kern-.08emT\kern-.1667em\lower.7ex\hbox{E}\kern-.125emX}}

\newcommand{\myalert}[2]{\ifthenelse{\boolean{disablecomment}}{#1}{\alert{#1} \todo{\small #2}}}

\begin{document}

\title{The Ingenuity Mars Helicopter Specified and Analyzed with the Real-time Mode-aware Dataflow Model}

\author{\IEEEauthorblockN{Guillaume Roumage$^\dagger$, Selma Azaiez$^\ddagger$, Cyril Faure$^\ddagger$, Stéphane Louise$^\ddagger$} \IEEEauthorblockA{$^\dagger$guillaume.roumage.research@proton.me \\ $^\dagger$$^\ddagger$\textit{Université Paris-Saclay, CEA, List, F-91120, Palaiseau, France} \\ $^\ddagger$firstname.lastname@cea.fr}}

\maketitle

\begin{abstract}
  Ingenuity is an autonomous Cyber-Pysical System (CPS) that has successfully completed more than 70 flights over Mars between 2021 and 2024. Ensuring the safety of its mission is paramount, as any failure could result in catastrophic economic damage and significant financial losses. Dataflow Models of Computation and Communication (DF MoCCs) serve as a formal framework for specifying and analyzing the timing behavior of such CPSs. In particular, the Real-time Mode-aware Dataflow (RMDF) model is highly suitable to specify and analyze real-time and mode-dependent \glspl{cps} like Ingenuity. This paper showcases the application of RMDF for the specification and analysis of Ingenuity. We propose a dataflow specification of Ingenuity, analyze its timing behavior, and provide a feasibility test. Finally, we proposed a plausible explanation of the timing anomaly that occurred during the sixth flight of Ingenuity.
\end{abstract}


\section{Introduction}

\glspl{cps} are reactive systems that detect environmental shifts through sensors, process this information using computational processes, and then use the output to control actuators. CPSs range from digital signal processing systems to embedded/cloud infrastructures, soft/hard real-time systems, and even a mix of all the above. These complex systems must operate reliably without threatening their internal processes. For example, a failure of the actuator in an autonomous car can lead to catastrophic consequences such as a car crash or a pedestrian accident.

The Ingenuity Mars helicopter is a real-time and autonomous CPS. It is a small coaxial helicopter designed by the \gls{nasa}. It has completed 72 flights over Mars from April 2021 to January 2024. Ingenuity is composed of sensors - an altimeter, two \glspl{imu}, and a camera - and actuators - eight motors. The sensors provide measurements to the navigation system, which estimates the state of Ingenuity. The control system then uses the state to keep Ingenuity as close as possible to a reference trajectory. A set of waypoints provides the reference trajectory. The control system sends commands to the motors to adjust the altitude, heading, and translation of Ingenuity.

The safety of Ingenuity is a critical issue. The safety of a CPS is the guarantee that the system will not endanger itself or its environment or lead to substantial financial losses. As an illustration of this latter point, the Ingenuity Mars mission cost raised \$85 million dollars~\cite{nasa_mars_2020}.

In this paper, we proposed to use \glspl{dfmocc} formalisms~\cite{roumage_survey_2022} to analyze the safety of Ingenuity. These formalisms permit the specification of a system with an oriented graph where actors represent computational units and channels represent communication links. \glspl{dfmocc} also provide static analyses to ensure the safety of the system's specification. For instance the \gls{rmdf} model is highly suitable to specify and analyze real-time and mode-dependent \glspl{cps} like Ingenuity. \gls{rmdf} can be used to guarantee, prior to execution, the existence of a memory-bounded execution (consistency), the existence of a deadlock-free execution (liveness), and to derive the timing behavior of Ingenuity.

\subsubsection{Contribution}

This paper presents the specification of the Ingenuity Mars helicopter with the dataflow model \gls{rmdf}~\cite{roumage_realtime_2025}. Besides guaranteeing the consistency and liveness of the system, \gls{rmdf} allows the derivation of the timing behavior of Ingenuity. A plausible explanation of the timing anomaly that occurred during the sixth flight of Ingenuity\footnote{https://science.nasa.gov/missions/mars-2020-perseverance/ingenuity-helicopter/surviving-an-in-flight-anomaly-what-happened-on-ingenuitys-sixth-flight/} is also given in this paper.

\subsubsection{Paper Organization}

This paper starts by presenting the background and terminology of the \gls{rmdf} model in \cref{sec:background}. As \gls{rmdf} extends \textsc{PolyGraph}~\cite{dubrulle_polygraph_2021}, which itself extends \gls{sdf} \cite{lee_synchronous_1987}, those two dataflow models are also presented. The Ingenuity mars helicopter use case is presented in \cref{sec:ingenuity} with its overview in \cref{sec:ingenuity-overview}, its RMDF specification in \cref{sec:ingenuity-rmdf-specification}, its timing behavior analysis in \cref{sec:ingenuity-timing-behavior-analysis}, and its feasibility test in \cref{sec:ingenuity-feasibility-test}. The timing anomaly of the sixth flight of Ingenuity is explained in \cref{sec:timing-anomaly}. Finally, \cref{sec:conclusion} concludes the paper.

\section{Background and Terminology}

\label{sec:background}

\gls{rmdf}~\cite{roumage_realtime_2025} is an extension of \textsc{PolyGraph}~\cite{dubrulle_polygraph_2021}, which is itself a superset of the \gls{sdf} model~\cite{lee_synchronous_1987}. In this section, we present \gls{sdf}, and then we explain how the \textsc{PolyGraph} model expands it.

\subsection{The SDF Model}

An \gls{sdf} specification is an oriented graph $G = (V, E)$ where $V$ is a finite set of \emph{actors} and $E$ is a finite set of \emph{channels}. An \gls{sdf} specification is characterized by a \emph{topology matrix} $G_\Gamma$ of size $|E| \times |V|$. An \emph{actor} is a computational unit that both produces and consumes data tokens every time it is executed. The atomic amount of data exchanged is known as a \emph{token}. The entry $(i, j)$ of $G_\Gamma$ is the number of tokens produced or consumed by the actor $v_j$ on the channel $c_i$ each time the actor $v_j$ is executed (this number is positive if the tokens are produced, and negative otherwise). An execution of an actor is called a \emph{job}.

An actor $v \in V$ has (optional) input and output ports connected to input and output channels of $E$. A channel $c_i = (v_j, v_k, n_{ij}, n_{ik}, init_i) \in E$ connects an output port of the actor $v_j \in V$ to an input port of the actor $v_k \in V$. This channel also has a production rate $n_{ij} \in \mathbb{N}^*$ (which is the entry $(i, j)$ of $G_\Gamma$), a consumption rate $n_{ik} \in \mathbb{N}^*$ (the entry $(i, k)$ of $G_\Gamma$), and a number of initial tokens $init_i \in \mathbb{N}$. An actor's job produces/consumes tokens to/from the buffer of its output/input channels according to the production/consumption rate. For the sake of simplicity in the rest of this paper, we denote $[c_i]$ the number of initial tokens of the channel $c_i$.

\subsection{The \textsc{PolyGraph} Model}

\subsubsection{Syntax and Semantic}

The \textsc{PolyGraph} model is a superset of the SDF model and expands it in two ways: the production and consumption rates become periodic sequences (as in CSDF~\cite{bilsen_cyclostatic_1996}), and a subset of actors may have timing constraints. The terminology of the static analysis of an SDF specification, i.e., consistency, liveness, consistent and live execution, and hyperperiod~\cite{lee_synchronous_1987}, is extended to a \textsc{PolyGraph} specification. Consistency ensures the system can be executed within a bounded memory, while liveness guarantees deadlock-free execution. A hyperperiod is a partially ordered set of actors' jobs that returns the system to its initial state within a given time frame.

\paragraph{Rational production and consumption rates}

Let $G = (V, E)$ be a \textsc{PolyGraph} specification and let $G_\Gamma = (\gamma_{ij}) \in \mathbb{Q}^{|E| \times |V|}$ be its topology matrix. The production and consumption rates are rational\footnote{A channel of a \textsc{PolyGraph} specification as defined in~\cite{dubrulle_polygraph_2021} has at least one integer rate. However, for the sake of simplicity, we consider in this paper that both rates are rational.}: a channel $c_i \in E$ is a tuple $(v_j, v_k, \gamma_{ij}, \gamma_{ik}, [c_i])$ such that $\gamma_{ij} \in \mathbb{Q}^*$ and $\gamma_{ik} \in \mathbb{Q}^*$, and $[c_i] \in \mathbb{Q}$ ($\mathbb{Q}^* = \mathbb{Q} \setminus \{ 0 \}$). Rational rates imply that the number of tokens produced/consumed can differ for each job. Specifically, a token is produced and stored in a channel if and only if sufficient fractional token parts are symbolically produced. Indeed, only an integer number of tokens may be produced or consumed. In other words, an actor with a rational production rate such as $1/n$ in the dataflow model actually produces data once every $n$ jobs during runtime, thus validating the precedence constraint toward the consumer once every $n$ of its jobs.
% These peculiarities create non-trivial cases when timing constraints (releases and deadlines) of actors without explicit real-time constraints in the specification are made explicit, as we will see in our contribution.

The rational production and consumption rates are natural consequences of the periodicity of real-time actors (with specified and imposed frequencies) together with the consistency of the system's model imposed by the topology matrix. The rational rates are a compact notation for a CSDF equivalency~\cite{bilsen_cyclostatic_1996}.

\paragraph{Timing constraints}

The actors of a \textsc{PolyGraph} specification may have a frequency constraint. Furthermore, they may also have a phase if they have a frequency constraint. The phase usually models the system's end-to-end latency. The frequency dictates the actor's execution at the specified rate, while the phase postpones its initial execution. Actors with frequency constraints are referred to as \emph{timed actors}. An execution of a consistent and live \textsc{PolyGraph} specification is an infinite repetition of its first hyperperiod. Although the number of jobs is not bounded during an execution, their timing constraints are cyclic over a hyperperiod.

\subsection{The RMDF model}

\subsubsection{Mode-Dependent Execution}

The \gls{rmdf} model is a superset of the \textsc{PolyGraph} model. Besides the ability to specify \glspl{cps} with relaxed real-time constraints, \gls{rmdf} can specify \glspl{cps} with mode-dependent execution. A mode-dependent \gls{cps} has a set of conditional execution branches. The choice of the execution branch to execute is usually based on runtime data. To illustrate, the \gls{rmdf} specification of \cref{fig:example-rmdf} has three conditional executions branches composed of actors $C_1, \dots, C_{n_1}$, $D_1, \dots, D_{n_2}$, and $E_1, \dots, E_{n_3}$.

\subsubsection{Splitters and Joiners}

The \gls{rmdf} model introduces two new actors: the \emph{splitter} and the \emph{joiner}. A \emph{splitter} is an actor that consumes a data token from its input channel and produces it on its output channels in the lexicographic order. A predefined number of tokens is sent to the first channel in the lexicographic order of the output channels, then to the second channel, and so on. The number of tokens sent to an output channel is the numerator of the production rate of that channel. An execution of a splitter consumed and produced one token at a time. A similar behavior holds for a \emph{joiner}, and the interested reader is referred to~\cite{roumage_realtime_2025} for more details.

\subsubsection{Mode Deciders, Controlled Splitters and Control Joiners}

The \gls{rmdf} model introduces three new actors: the \emph{mode decider}, the \emph{controlled splitter}, and the \emph{controlled joiner}. The \emph{mode decider} is responsible for deciding the execution branch to execute (actor $B$ of \cref{fig:example-rmdf} is a mode decider). A \emph{controlled splitter} and a \emph{controlled joiner} are used to route data tokens to/from the execution branches. The \emph{mode decider} is connected to the \emph{controlled splitter} and the \emph{controlled joiner} through a \emph{control channel} that carries a \emph{control token}. This control token is used to decide from/to which channel tokens are consumed/produced by assigning a value to parametric rates. In \cref{fig:example-rmdf}, those parametric rates are $m_1$, $m_2$ and $m_3$. In other words, the \emph{controlled splitter} and the \emph{controlled joiner} have an \emph{switch-case} structure that decides how a data token is produced and consumed.

\subsubsection{Control areas}

\begin{figure*}[htbp]
  \centering
  \begin{tikzpicture}
    \node[actor] (a) at (-1,-2.5) {A \\ \emph{100 Hz}};
    \node[virtualactor] (dup-1) at (-1,0) {Duplicater 1};
    \node[virtualactor] (dup-2) at (7.5,-2.5) {Duplicater 2};
    \node[virtualactor] (cs) at (3,0) {Controlled \\ splitter};
    \node[virtualactor] (cj) at (12,0) {Controlled \\ joiner};
    \node[decider] (b) at (2,-2.5) {B};
    \node[actor] (c1) at (6.75,1) {C$_1$};
    \node (ci) at (7.5,1) {\dots};
    \node[actor] (cl) at (8.25,1) {C$_{n_1}$};
    \node[actor] (d1) at (6.75,0) {D$_1$};
    \node (di) at (7.5,0) {\dots};
    \node[actor] (dm) at (8.25,0) {D$_{n_2}$};
    \node[actor] (e1) at (6.75,-1) {E$_1$};
    \node (ei) at (7.5,-1) {\dots};
    \node[actor] (en) at (8.25,-1) {E$_{n_3}$};
    \node[actor] (f) at (12,-2.5) {F \\ \emph{100 Hz}};
    \draw[arc] (a) to (dup-1);
    \draw[arc] (cj) to (f);
    \draw[arc] (dup-1) to (cs);
    \draw[arc] (dup-1) to (b);
    \draw[arc] (cs) to node[sloped,fill=white,pos=0.35] {$m_1$} (c1);
    \draw[arc] (cs) to node[sloped,fill=white,pos=0.35] {$m_2$} (d1);
    \draw[arc] (cs) to node[sloped,fill=white,pos=0.35] {$m_3$} (e1);
    \draw[arc] (en) to node[sloped,fill=white,pos=0.65] {$m_3$} (cj);
    \draw[arc] (dm) to node[sloped,fill=white,pos=0.65] {$m_2$} (cj);
    \draw[arc] (cl) to node[sloped,fill=white,pos=0.65] {$m_1$} (cj);
    \draw[arc,dashed] (b) to (dup-2);
    \draw[arc,dashed] (dup-2) to (cs);
    \draw[arc,dashed] (dup-2) to (cj);
    \draw[arc] (-2,-4) to (-1,-4);
    \node[anchor=west] at (-1,-4) {Data channel};
    \draw[arc,dashed] (-2,-5) to (-1,-5);
    \node[anchor=west] at (-1,-5) {Control channel};
    \node[actor] (toto) at (3.5,-4) {\textcolor{white}{A}};
    \node[anchor=west] at (toto.east) {Usual actor};
    \node[decider, aspect=0.5] (titi) at (3.5,-6) {\textcolor{white}{A}};
    \node[anchor=west] at (titi.east) {Mode decider};
    \node[virtualactor] (tutu) at (3.5,-5) {\textcolor{white}{A}};
    \node[anchor=west] at (tutu.east) {Routing actor};
    \node[anchor=west,align=left] at (7,-5) {$M = \{ M_1, M_2, M_3 \} $ \\ $M_1(m_1) = 1, M_2(m_2) = 0, M_3(m_3) = 0$ \\ $M_1(m_1) = 0, M_2(m_2) = 1, M_3(m_3) = 0$ \\ $M_1(m_1) = 0, M_2(m_2) = 0, M_3(m_3) = 1$};
  \end{tikzpicture}
  \caption{An example of an RMDF specification. A usual actor is an actor which is neither (non-) controlled splitter or joiner, nor a mode decider.}
  \label{fig:example-rmdf}
\end{figure*}
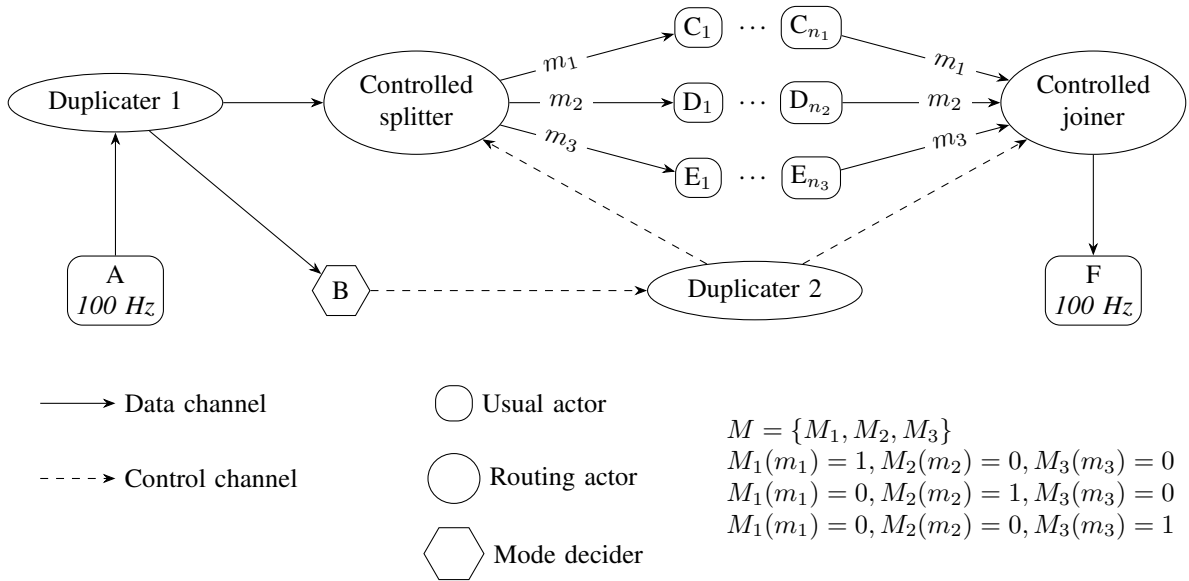

The RMDF model introduces the concept of \emph{control areas}. A control area is a set of actors that are conditioned by a \emph{mode decider}. The control area of the mode decider $B$ of \cref{fig:example-rmdf} is composed of actors $C_1, \dots, C_{n_1}$, $D_1, \dots, D_{n_2}$, and $E_1, \dots, E_{n_3}$. In order to ensure the static analysis of an RMDF specification, restrictions are applied on control areas. Those restrictions are detailed in Property 1 of~\cite{roumage_realtime_2025}.

\section{The Ingenuity Mars Helicopter}

\label{sec:ingenuity}

The main contribution of this paper is the specification of the Ingenuity Mars helicopter with the RMDF model. \cref{sec:ingenuity-overview} provides an overview of Ingenuity, \cref{sec:ingenuity-rmdf-specification} presents the RMDF specification of Ingenuity, \cref{sec:ingenuity-timing-behavior-analysis} presents the timing behavior analysis of Ingenuity and \cref{sec:ingenuity-feasibility-test} presents a feasibility test for Ingenuity.

\subsection{Overview of Ingenuity}

\label{sec:ingenuity-overview}

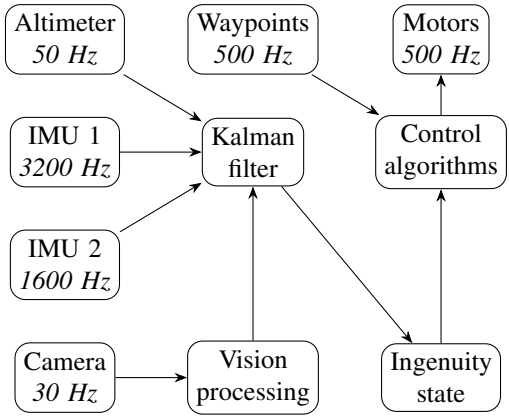
\begin{figure}[htbp]
  \centering
  \begin{tikzpicture}
    \node[actor] (alt) at (0,0) {Altimeter \\ \emph{50 Hz}};
    \node[actor] (imu1) at (0, -1.5) {IMU 1 \\ \emph{3200 Hz}};
    \node[actor] (imu2) at (0, -3) {IMU 2 \\ \emph{1600 Hz}};
    \node[actor] (cam) at (0, -4.5) {Camera \\ \emph{30 Hz}};
    \node[actor] (vis) at (2.5, -4.5) {Vision \\ processing};
    \node[actor] (kal) at (2.5, -1.5) {Kalman \\ filter};
    \node[actor] (ing) at (5, -4.5) {Ingenuity \\ state};
    \node[actor] (ctrl) at (5, -1.5) {Control \\ algorithms};
    \node[actor] (mot) at (5, 0) {Motors \\ \emph{500 Hz}};
    \node[actor] (way) at (2.5, 0) {Waypoints \\ \emph{500 Hz}};
    \draw[arc] (alt) to (kal);
    \draw[arc] (imu1) to (kal);
    \draw[arc] (imu2) to (kal);
    \draw[arc] (kal) to (ing);
    \draw[arc] (ing) to (ctrl);
    \draw[arc] (ctrl) to (mot);
    \draw[arc] (cam) to (vis);
    \draw[arc] (vis) to (kal);
    \draw[arc] (way) to (ctrl);
  \end{tikzpicture}
  \caption[An overview of the Ingenuity helicopter]{An overview of the Ingenuity helicopter.}
  \label{fig:ingenuity_fusion_system}
\end{figure}

Ingenuity operates independently due to communications delays of at least 5 minutes between Earth and Mars\footnote{https://mars.nasa.gov/mars2020/spacecraft/rover/communications/}. Hence, real-time navigation from Earth is impossible. Ingenuity uses a Kalman filter~\cite{kalman_new_1960} to fuse measurements from three sensors (cf. \cref{fig:ingenuity_fusion_system}, with a redundancy of the \gls{imu}) and regularly updates its \emph{state}, i.e.,  its position, velocity, attitude, and its angular rate. Ingenuity has two IMUs for redundancy, which measure 3-axis acceleration at 1600 Hz and angular rates at 3200 Hz, and an inclinometer is used before flights to calibrate IMU biases. Additionally, an altimeter measures the distance to the ground at 50 Hz, and a downward-looking navigation camera provides images at 30 Hz~\cite{balaram_mars_2018}. Ingenuity's actuators consist of eight motors - four for each rotor. One motor keeps the rotation speed constant, while the remaining motors control the blade pitch, affecting Ingenuity's motion.

\subsubsection{Vision-Based Navigation}

Ingenuity uses a \emph{dead reckoning} process to compute its current state: this navigation technique estimates the current vehicle's state based on previous sensor measurements. However, the dead reckoning process is susceptible to drift. To counter this issue, Ingenuity uses a vision-based navigation system. Periodically, it captures images of the Martian ground, detects visual features, and tracks them from one image to the next. Some images are regularly classified as \emph{base frames}, and the features in these images become the \emph{pseudo landmarks}, while other images are classified as \emph{search frames} whose features will be computed and compared to the pseudo landmarks. Therefore, Ingenuity can determine a visual shift and correct the drift in the dead reckoning process by comparing the current features with the most recent pseudo landmarks. A new base frame is generated if certain conditions are not met, including reaching the maximum number of search frames between two consecutive base frames or a drop in the number of tracked features (tracked features are features from a search frame that are successfully matched with pseudo landmarks)~\cite{bayard_visionbased_2019}.

\subsubsection{Navigation Algorithm}

The IMU measurements undergo pre-processing with filters, followed by integration and downsampling from 3200 Hz and 1600 Hz to 500 Hz. The downsampled IMU measurements are corrected with IMU biases computed from the previous state estimation, then sent to the navigation filter component and buffered for future use by a state propagation process. The navigation filter operates as a Kalman filter and fuses the IMU data, altimeter measurements, and vision system results to estimate Ingenuity's state. The resulting state is sent back to the state propagation process, which propagates the state, i.e., predicts the state for the next time steps, using the buffered IMU data. The control system of Ingenuity is responsible for keeping the helicopter as close as possible to a reference trajectory, given as a set of waypoints: each of them specifies an (x,y)-position, a ground-relative altitude, a translation heading, a dwell heading, and a dwell time. Ingenuity advances between waypoints by performing a sequence of maneuvers: (1) altitude adjustment, (2) turn to specified flight heading, (3) horizontal translation, (4) turn to dwell heading, (5) pausing for the specified dwell time~\cite{grip_flight_2019}.

\subsection{An RMDF Specification of Ingenuity}

\label{sec:ingenuity-rmdf-specification}

We have translated the textual specifications of Ingenuity~\cite{sanmartin_minimal_2017,balaram_mars_2018,grip_guidance_2018,bayard_visionbased_2019,grip_flight_2019,grip_flying_2022} to an RMDF specification depicted in \cref{fig:ingenuity-rmdf-model}. Actors \emph{IMU}, \emph{Altimeter}, and \emph{Camera} are the three sensors of Ingenuity, with frequencies of 500 Hz, 50 Hz, and 30 Hz, respectively. Ingenuity downsamples its \emph{IMU} data from 3200 Hz and 1600 Hz to 500 Hz, so we assume this downsampling is already done. The specification of Ingenuity of \cref{fig:ingenuity-rmdf-model} is consistent and live, as verified by our internal scripts implementing the static analyses of RMDF specifications~\cite{roumage_realtime_2025}.

\subsubsection{Vision System}

The vision system is depicted in the top tier of \cref{fig:ingenuity-rmdf-model}. The actors from \emph{Camera} to \emph{Feature Match} belong to the vision system of Ingenuity. The \emph{Laber Decider} is responsible for classifying a frame as a base or search frame. In the real system, a new base frame is triggered if at least one of the following conditions is not met~\cite{bayard_visionbased_2019}: (1) the number of tracked features is below a threshold, (2) the number of search frames since the last base frame is above a threshold, (3) visual features are not well-distributed. Note that those conditions cannot be predetermined at design time as they depend on the frame's environment and quality. RMDF allows such dynamic behavior to be specified.

The decided mode is sent to the control port of the \emph{Controlled splitter} and the \emph{Controlled joiner} through \emph{Duplicater 2}. \emph{Controlled splitter} and \emph{Controlled joiner} enclose an \emph{if-then-else} structure that decides how a frame is processed. In the \emph{then} branch, the features of a search frame are tracked by \emph{Feature Tracking} before being filtered with a \emph{Filtering Procedure}. In the \emph{else} branch, the features of a base frame are used to update the \emph{Pseudo Landmarks}. 

There are two conditional branches: \emph{Feature Tracking} $\to$ \emph{Filtering Procedure} and \emph{Pseudo Landmarks}. The parametric rates $m_1$ and $m_2$ on channels $c_2, c_3, c_4$ and $c_5$ determine whether or not the conditional branch is taken during an iteration. The \emph{Controlled splitter} and \emph{Controlled joiner} use the control token to decide from/to which channel tokens are consumed/produced. Finally, the self-loop on \emph{Feature Tracking} models the continuous tracking of features from one search frame to the next.

\subsubsection{Navigation System}

The actors from \emph{IMU} and \emph{Altimeter} to \emph{State Propagation} belong to Ingenuity's navigation system. The \emph{Navigation Filter} inherits the frequency of 500 Hz from the \emph{IMU}. Thus, it has to oversample the vision processing result with a consumption rate of $3/50$ as the frequency of the \emph{Camera} is 30 Hz. The same logic applies to the channel \emph{Duplicater~4} $\to$ \emph{Navigation Filter} and \emph{Duplicater~4} $\to$ \emph{State Propagation}.

The following paragraph details the origin of the initial token of $1/50$ on the channel $c_6$. \emph{Motors} operates at a frequency of 500 Hz (i.e., 2 ms = $6 \cdot (1/3)$ ms) with a phase of 1 ms. As a result of the liveness analysis, a tick in the symbolic execution of the Ingenuity specification is equivalent to a time advance of $1/3$ ms. Consequently, \emph{Motors} is required to execute at tick $\tau$ if $\tau \equiv 3 \mod{6}$. At tick $99$, \emph{Motors} must execute for the 17-th time, and \emph{Navigation Filter} must also execute for the 17-th time due to data dependencies. Therefore, $17 \cdot (3/50)$ tokens are required on $c_6$. However, at the 99-th tick, \emph{Camera} has executed only once. Its second execution occurs later, at the 100-th tick. Thus, there is only one token on $c_6$, which is not enough for the 17-th execution of \emph{Navigation Filter} and a deadlock occur. The initial token of $1/50$ on $c_6$ avoids this deadlock by enabling the 17-th execution of \emph{Navigation Filter}.

The \emph{Navigation Filter} is the core of the navigation system of Ingenuity: it fuses the measurements from the sensors to estimate the state of Ingenuity. Then, \emph{State Propagation} is responsible for predicting the state of Ingenuity for the next time steps. Note that \emph{IMU-Correction} corrects the \emph{IMU} measurements with the biases computed from the previous state estimation. The integration of the \emph{IMU} data is redundant with \emph{NC - IMU - Integration} and \emph{FC - IMU - Integration}. The integration is performed in two hardware units, denoted as the \emph{Navigation Computer} and \emph{Flight Computer} in Ingenuity's specifications. We do not detail the difference between these two units here; the interested reader is refered to~\cite{grip_flight_2019}. An initial token on the channel $c_9$ models the initial bias correction. An inclinometer initializes the IMU biases before the flight in the real system.

\subsubsection{Control System}

The actors \emph{Control Altitude}, \emph{Control Yaw 1}, \emph{Control Translation}, and \emph{Control Yaw 2} are responsible for triggering Ingenuity's motors to keep it as close as possible to the reference trajectory. This latter is provided by a set of waypoints produced by \emph{Waypoints}. The four control actors compare the current state of Ingenuity with the expected state and send the appropriate commands to the motors. \cref{tab:ingenuity-channels-specification} provides details on the production and consumption rate specifications. They are chosen such that, during an execution, \emph{Control Altitude} is first executed, then \emph{Control Yaw 1} adjusts the heading, followed by the \emph{Control Translation}, and finally, another heading adjustment with \emph{Control Yaw 2}. Then, the cycle repeats.

\begin{figure*}[htbp]
  \begin{subfigure}[c]{\textwidth}
    \centering
    \caption{An RMDF specification of Ingenuity deployed after its sixth flight. Rates of $1$ and initial tokens of $0$ are omitted for clarity.}
    \label{fig:ingenuity-specification}
    \begin{tikzpicture}
      \node[actor] (cam) at (0,-5.5) {\underline{Cam}era \\ (Cam) \emph{30 Hz}};
      \node[actor] (fd) at (-4,-5.5) {\underline{F}eature\\\underline{D}etection (FD)};
      \node[virtualactor] (dup1) at (-4,-7) {Duplicater 1};
      \node[decider] (ld) at (-4,-9) {\underline{L}abel\\\underline{D}ecider\\(LD)};  
      \node[virtualactor] (dup2) at (0,-9) {Duplicater 2};
      \node[virtualactor] (cs) at (0,-7) {Controlled\\splitter};
      \node[actor] (ft) at (5,-5.5) {\underline{F}eature\\\underline{T}racking (FT)};
      \node[actor] (fp) at (8,-7) {\underline{F}iltering\\\underline{P}rocedure (FP)};
      \node[actor] (pl) at (5,-7) {\underline{P}seudo\\\underline{L}andmarks (PL)};
      \node[actor] (fm) at (8,-9) {\underline{F}eature\\\underline{M}atch (FM)};
      \node[virtualactor] (cj) at (3.5,-9) {Controlled\\joiner};
      \draw[arc] (cam) to (fd);
      \draw[arc] (fd) to (dup1);
      \draw[arc] (dup1) to (ld);
      \draw[arccontrol] (ld) to (dup2);
      \draw[arccontrol] (dup2) to (cs);
      \draw[arccontrol] (dup2) to (cj);
      \draw[arc] (cs) to node[pos=0.5,fill=white] {$c_2$} (ft);
      \draw[arc] (cs) to node[pos=0.5,fill=white] {$c_3$} (pl);
      \draw[arc] (pl) to node[pos=0.5,fill=white] {$c_5$} (cj);
      \draw[arc] (ft) to (fp);
      \draw[arc] (ft.45) to[out=90,in=90] node[pos=0.5,fill=white] {$c_1$} (ft.135);
      \draw[arc] (fp) to node[pos=0.5,fill=white] {$c_4$} (cj);
      \draw[arc] (cj) to (fm);
      \draw[arc] (dup1) to (cs);
      \node[actor] (imu) at (-4,-11) {\underline{IMU}\\\emph{500 Hz}};
      \node[actor] (ic) at (0,-11) {\underline{IMU}-\underline{C}orrection\\(IMU-CORR)};
      \node[actor] (nf) at (5,-11) {\underline{N}avigation\\\underline{F}ilter (NF)};
      \node[actor] (alt) at (8,-11) {\underline{Alt}imeter\\(ALT) \emph{50 Hz}};
      \node[virtualactor] (dup3) at (-4,-12.5) {Duplicater 3};
      \node[actor] (nc-imu-int) at (2,-12.5) {\underline{NC} - \underline{IMU}\\\underline{Int}egration\\(NC-IMU-INT)};
      \node[actor] (fc-imu-int) at (-4,-14) {\underline{FC} - \underline{IMU}\\\underline{Int}egration\\(FC-IMU-INT)};
      \node[actor] (sp) at (5,-14) {\underline{S}tate \\ \underline{P}ropagation (SP)};
      \node[virtualactor] (dup4) at (8,-12.5) {Duplicater 4};
      \draw[arc] (imu) to (ic);
      \draw[arc] (ic) to (dup3);
      \draw[arc] (dup3) to (nc-imu-int);
      \draw[arc] (dup3) to (fc-imu-int);
      \draw[arc] (nc-imu-int) to (nf);
      \draw[arc] (fc-imu-int) to (sp);
      \draw[arc] (nf) to (sp);
      \draw[arc] (nf) to node[pos=0.5,fill=white] {$c_9$} (ic);
      \draw[arc] (alt) to (dup4);
      \draw[arc] (dup4) to node[pos=0.5,fill=white] {$c_8$} (sp);
      \draw[arc] (dup4) to node[pos=0.5,fill=white] {$c_7$} (nf);
      \draw[arc] (fm) to node[pos=0.5,fill=white] {$c_6$} (nf);
      \node[virtualactor] (spl1) at (-4,-16) {Splitter 1};
      \node[actor] (ctrl-alt) at (2,-16) {\underline{C}ontrol\\\underline{A}ltitude (CA)};
      \node[actor] (ctrl-yaw-1) at (2,-17.5) {\underline{C}ontrol\\\underline{Y}aw \underline{1} (CY1)};
      \node[actor] (ctrl-trans) at (2,-19) {\underline{C}ontrol\\\underline{T}ranslation (CT)};
      \node[actor] (ctrl-yaw-2) at (2,-20.5) {\underline{C}ontrol\\\underline{Y}aw \underline{2} (CY2)};
      \node[virtualactor] (spl2) at (-4,-20.5) {Splitter 2};
      \node[virtualactor] (joiner) at (8,-17.5) {Joiner};
      \node[actor] (waypoints) at (-4,-18.25) {\underline{Way}points\\(WAY) \emph{500 Hz}};
      \node[actor] (mot) at (8,-20.5) {\underline{Mot}ors\ (MOT) \\ \emph{500 Hz}, \emph{+ 1 ms}};
      \draw[arc] (sp) to (spl1);
      \draw[arc] (spl1) to node[pos=0.3,fill=white] {$c_{10}$} (ctrl-alt);
      \draw[arc] (spl1) to node[pos=0.3,fill=white] {$c_{11}$} (ctrl-yaw-1);
      \draw[arc] (spl1) to node[pos=0.3,fill=white] {$c_{12}$} (ctrl-trans);
      \draw[arc] (spl1) to node[pos=0.3,fill=white] {$c_{13}$} (ctrl-yaw-2);
      \draw[arc] (spl2) to node[pos=0.3,fill=white] {$c_{14}$} (ctrl-alt);
      \draw[arc] (spl2) to node[pos=0.3,fill=white] {$c_{15}$} (ctrl-yaw-1);
      \draw[arc] (spl2) to node[pos=0.3,fill=white] {$c_{16}$} (ctrl-trans);
      \draw[arc] (spl2) to node[pos=0.3,fill=white] {$c_{17}$} (ctrl-yaw-2);
      \draw[arc] (waypoints) to (spl2);
      \draw[arc] (ctrl-alt) to node[pos=0.7,fill=white] {$c_{18}$} (joiner);
      \draw[arc] (ctrl-yaw-1) to node[pos=0.7,fill=white] {$c_{19}$} (joiner);
      \draw[arc] (ctrl-trans) to node[pos=0.7,fill=white] {$c_{20}$} (joiner);
      \draw[arc] (ctrl-yaw-2) to node[pos=0.7,fill=white] {$c_{21}$} (joiner);
      \draw[arc] (joiner) to (mot);
      \node[rotate=90] at (-6,-7) {Vision system};
      \draw[dashed] (-6,-10.25) -- (9,-10.25);
      \node[rotate=90] at (-6,-12.5) {Navigation system};
      \draw[dashed] (-6,-15) -- (9,-15);
      \node[rotate=90] at (-6,-18) {Control system};
    \end{tikzpicture}
  \end{subfigure}
  \begin{subfigure}[c]{\textwidth}
    \centering
    \caption{Channels' specifications of the above RMDF specification.}
    \footnotesize
    \resizebox{\textwidth}{!}{
      \begin{tabular}{|c|c|c|c||c|c|c|c||c|c|c|c|}
        \toprule
        chan. & cons. & prod. & init. & chan. & cons. & prod. & init. & chan. & cons. & prod. & init. \\
        \midrule
        $c_1$ & 1 & 1 & 1 & $c_8$ & 1 & $1/10$ & 0 & $c_{15}$ & $1/4$ & 1 & 0 \\
        \midrule
        $c_2$ & $m_1$ & 1 & 0 & $c_9$ & 1 & 1 & $1$ & $c_{16}$ & $1/4$ & 1 & 0 \\
        \midrule
        $c_3$ & $m_2$ & 1 & 0 & $c_{10}$ & $1/4$ & 1 & 0 & $ c_{17}$ & $1/4$ & 1 & 0 \\
        \midrule
        $c_4$ & 1 & $m_1$ & 0 & $c_{11}$ & $1/4$ & 1 & 0 & $c_{18}$ & 1 & $1/4$ & 0 \\
        \midrule
        $c_5$ & 1 & $m_2$ & 0 & $c_{12}$ & $1/4$ & 1 & 0 & $c_{19}$ & 1 & $1/4$ & 0 \\
        \midrule
        $c_6$ & $3/50$ & 1 & $1/50$ & $c_{13}$ & $1/4$ & 1 & 0 & $c_{20}$ & 1 & $1/4$ & 0 \\
        \midrule
        $c_7$ & 1 & $1/10$ & 0 & $c_{14}$ & $1/4$ & 1 & 0 & $c_{21}$ & 1 & $1/4$ & 0 \\
        \bottomrule
      \end{tabular}
    }
    \label{tab:ingenuity-channels-specification}
  \end{subfigure}
  \caption{An RMDF specification of Ingenuity deployed after its sixth flight and the specification of its channel characteristics.}
  \label{fig:ingenuity-rmdf-model}
\end{figure*}

\subsubsection{Fidelity of the Ingenuity Specification}

In the vision system, we have simplified the labeling process of frames. In the real system, a frame can be classified as both a base frame and a search frame, e.g., when the number of tracked features is below a threshold. In such cases, the current features replace the pseudo landmarks, causing a frame initially classified as a search frame to be also classified as a base frame. Our specification operates on the assumption that a frame can only be a base or search frame, but never both simultaneously. As a result, continuous tracking of features from a search frame to a base frame is not supported by our specification. As a future work, this could be improved by adding another mode that would allow a frame to be sent to multiple conditional branches.

The navigation system assumes that the IMU data are already downsampled to 500 Hz. However, in the real system, the IMU data are downsampled from 3200 Hz and 1600 Hz to 500 Hz. In addition, we simplify by assuming that the initial bias correction is already incorporated. In the real system, the inclinometer initializes the IMU biases before flight, and this initialization is represented by the initial token on $c_9$ in \cref{fig:ingenuity-specification}.

In our specification, we implicitly operate under the assumption that movement during each maneuver (altitude, yaw, and translation) occurs almost instantaneously, although this does not accurately reflect real-world conditions. After performing the \emph{Control Yaw 2} maneuver, Ingenuity also undergoes a specified wait time. Our model does not account for this waiting time. Furthermore, based on the specification, we know that the end-to-end latency of Ingenuity is 10 ms, which implies that \emph{Motors} phase should be 10 ms. However, the RMDF model imposes a limitation where the phase must be strictly less than the period (as for the \textsc{PolyGraph} model). Due to this constraint, we have set the \emph{Motors} phase to the maximum allowable value of 1 ms. It would be interesting to investigate how to relax this constraint in future work.

Finally, Ingenuity has different phases of flight, including \emph{climb}, \emph{waypoint tracking}, and \emph{descent}, as mentioned in \cref{tab:phases-of-flight}. Each phase of flight involves turning on and off different sensors. Specifically, the \emph{Camera} and \emph{Altimeter} are disabled when Ingenuity is under 1-meter altitude to avoid erroneous measurements due to dust kickup from rotor downwash. Altitude thresholds initiate the transition from one phase to another. Ingenuity also has a fault response mode, a \enquote{land as soon as possible} mode. In this mode, the \emph{Camera} is disabled, while the \emph{IMU} and \emph{Altimeter} remain enabled. The transition to the fault response mode is triggered when 50 consecutive packets containing vision processing results fail to arrive at \emph{State Propagation}~\cite{grip_flight_2019}. Our specification only considers the \emph{Waypoints Tracking} phase, and there is no fault response mode.

\begin{table}[htbp]
  \centering
  \caption{Phases of flight of Ingenuity.}
  \label{tab:phases-of-flight}
  \resizebox{\columnwidth}{!}{
    \begin{threeparttable}[htbp]
      \begin{tabular}{|c||c|c|c||c|}
        \toprule
        \makecell{\textbf{Phase of} \\ \textbf{flight}} & \textbf{Climb}\tnote{1} & \makecell{\textbf{Waypoints} \\ \textbf{Tracking}} & \textbf{Descent}\tnote{2} & \makecell{\textbf{Fault Response} \\ \textbf{Mode}} \\
        \midrule
        \makecell{Altitute \\ threshold} & $<$ 1 meter & $\geq$ 1 meter & $<$ 1 meter & any \\ 
        \midrule
        Camera & $\circ$ & $\bullet$ & $\circ$ & $\circ$ \\
        \midrule
        Altimeter & $\circ$ & $\bullet$ & $\circ$ & $\bullet$ \\
        \midrule
        IMU & $\bullet$ & $\bullet$ & $\bullet$ & $\bullet$ \\
        \bottomrule
      \end{tabular}
      \begin{tablenotes}
        \item[1] \emph{climb} contains also a \emph{take off} phase where control of the helicopter is limited
        \item[2] a monitoring of the vertical velocity is also enabled when altitude is less than 0.5 meters
      \end{tablenotes}
    \end{threeparttable}
  }
\end{table}

\subsection{Timing Behavior Analysis of the RMDF Specification of Ingenuity}

\label{sec:ingenuity-timing-behavior-analysis}

The specification of Ingenuity presented in this paper is consistent and live, i.e., it has a memory-bounded and a deadlock-free execution. Those properties are verified using internal tools of the RMDF library. The timing propagation presented in~\cite{roumage_static_2024} has been applied to the Ingenuity specification and result are presented in \cref{tab:ingenuity-timing-constraints}, assuming \glspl{bcet} and \glspl{wcet} are 0.12 and 0.20 ms, respectively, for all actors except \emph{Controlled Splitter} and \emph{Controlled Joiner} which have a constant execution time of 0 ms. 

\begin{table*}[p]
  \centering
  \caption{Releases, deadlines and execution windows length of the $n$-th job for each actor of Ingenuity, assuming \gls{bcet} and \gls{wcet} of all actors are 0.12 and 0.20 ms, respectively, except for \emph{Controlled Splitter} and \emph{Controlled Joiner} which have a constant execution time of 0 ms.}
  \label{tab:ingenuity-timing-constraints}
  \resizebox{\textwidth}{!}{
    \begin{tabular}{|c|c|c|c|}
      \toprule
      \multirow{2}{*}{\textbf{Actor}} & \multicolumn{3}{c|}{\textbf{Timing constraints of the $n$-th job}} \\
      \cmidrule{2-4}
      & \textbf{Release} & \textbf{Deadline} & \textbf{Exec. Windows} \\
      \midrule
      \glsentryshort{cam} & $(100 \cdot (n-1))/3$ & $\begin{cases} 6/5 + \lfloor (100 \cdot (n-1)) \rfloor & \text{ if } (n-1) \mod{3} = 0 \\ 176/5 + \lfloor 100 \cdot (n-1) \rfloor & \text{ if } (n-1) \mod{3} = 1 \\ 336/5 + \lfloor 100 \cdot (n-1) \rfloor & \text{ if } (n-1) \mod{3} = 2 \end{cases}$ & $\begin{cases} 6/5 & = 1.20~ms\\25/15 & \simeq 1.87~ms\\8/15 & \simeq 0.53~ms \end{cases}$ \\
      \midrule
      \glsentryshort{fd} & $(9 + 2500 \cdot (n-1))/75$ & $\begin{cases} 7/5 + \lfloor 100 \cdot (n-1) \rfloor & \text{ if } (n-1) \mod{3} = 0 \\ 177/5 + \lfloor 100 \cdot (n-1) \rfloor & \text{ if } (n-1) \mod{3} = 1 \\ 337/5 + \lfloor 100 \cdot (n-1) \rfloor & \text{ if } (n-1) \mod{3} = 2 \end{cases}$ & $\begin{cases} 32/25 & = 1.96~ms\\146/75 & \simeq 1.94~ms\\46/75 & \simeq 0.61~ms \end{cases}$ \\
      \midrule
      \glsentryshort{ld} & $(18 + 2500 \cdot (n-1))/75$ & $\begin{cases} 8/5 + \lfloor 100 \cdot (n-1) \rfloor & \text{ if } (n-1) \mod{3} = 0 \\ 178/5 + \lfloor 100 \cdot (n-1) \rfloor & \text{ if } (n-1) \mod{3} = 1 \\ 338/5 + \lfloor 100 \cdot (n-1) \rfloor & \text{ if } (n-1) \mod{3} = 2 \end{cases}$ & $\begin{cases} 34/25 & = 1.36~ms\\152/75 & \simeq 2.03~ms\\52/75 & \simeq 0.70~ms \end{cases}$ \\
      \midrule
      CS & $(27 + 2500 \cdot (n-1))/75$ & $\begin{cases} 8/5 + \lfloor 100 \cdot (n-1) \rfloor & \text{ if } (n-1) \mod{3} = 0 \\ 178/5 + \lfloor 100 \cdot (n-1) \rfloor & \text{ if } (n-1) \mod{3} = 1 \\ 338/5 + \lfloor 100 \cdot (n-1) \rfloor & \text{ if } (n-1) \mod{3} = 2 \end{cases}$ & $\begin{cases} 31/25 & = 1.24~ms\\143/75 & \simeq 1.91~ms\\43/75 & \simeq 0.57~ms \end{cases}$ \\
      \midrule
      \glsentryshort{ft} & $(27 + 2500 \cdot (n-1))/75$ & $\begin{cases} 9/5 + \lfloor 100 \cdot (n-1) \rfloor & \text{ if } (n-1) \mod{3} = 0 \\ 179/5 + \lfloor 100 \cdot (n-1) \rfloor & \text{ if } (n-1) \mod{3} = 1 \\ 339/5 + \lfloor 100 \cdot (n-1) \rfloor & \text{ if } (n-1) \mod{3} = 2 \end{cases}$ & $\begin{cases} 36/25 & = 1.44~ms\\158/75 & \simeq 2.11~ms\\58/75 & \simeq 0.78~ms \end{cases}$ \\
      \midrule
      \glsentryshort{fp} & $(36 + 2500 \cdot (n-1))/75$ & $\begin{cases} 2 + \lfloor (100 \cdot (n-1))/3 \rfloor & \text{ if } (n-1) \mod{3} = 0 \\ 36 + \lfloor (100 \cdot (n-1))/3 \rfloor & \text{ if } (n-1) \mod{3} = 1 \\ 68 + \lfloor (100 \cdot (n-1))/3 \rfloor & \text{ if } (n-1) \mod{3} = 2 \end{cases}$ & $\begin{cases} 38/25 & = 1.52~ms\\164/75 & \simeq 2.19~ms\\64/75 & \simeq 0.85~ms \end{cases}$ \\
      \midrule
      \glsentryshort{pl} & $(27 + 2500 \cdot (n-1))/75$ & $\begin{cases} 2 + \lfloor (100 \cdot (n-1))/3 \rfloor & \text{ if } (n-1) \mod{3} = 0 \\ 36 + \lfloor (100 \cdot (n-1))/3 \rfloor & \text{ if } (n-1) \mod{3} = 1 \\ 68 + \lfloor (100 \cdot (n-1))/3 \rfloor & \text{ if } (n-1) \mod{3} = 2 \end{cases}$ & $\begin{cases} 41/25 & = 1.64~ms\\173/75 & \simeq 2.31~ms\\73/75 & \simeq 0.97~ms \end{cases}$ \\
      \midrule
      CJ & $(9 + 500 \cdot (n-1))/15$ & $\begin{cases} 2 + \lfloor (100 \cdot (n-1)) \rfloor & \text{ if } (n-1) \mod{3} = 0 \\ 36 + \lfloor (100 \cdot (n-1)) \rfloor & \text{ if } (n-1) \mod{3} = 1 \\ 68 + \lfloor (100 \cdot (n-1)) \rfloor & \text{ if } (n-1) \mod{3} = 2 \end{cases}$ & $\begin{cases} 7/5 & = 1.40~ms\\31/15 & \simeq 2.07~ms\\11/15 & \simeq 0.73~ms \end{cases}$ \\
      \midrule
      \glsentryshort{fm} & $(9 + 500 \cdot (n-1))/15$ & $\begin{cases} 11/5 + \lfloor 100 \cdot (n-1) \rfloor & \text{ if } (n-1) \mod{3} = 0 \\ 181/5 + \lfloor 100 \cdot (n-1) \rfloor & \text{ if } (n-1) \mod{3} = 1 \\ 341/5 + \lfloor 100 \cdot (n-1) \rfloor & \text{ if } (n-1) \mod{3} = 2 \end{cases}$ & $\begin{cases} 8/5 & = 1.60~ms\\34/15 & \simeq 2.27~ms\\14/15 & \simeq 0.93~ms \end{cases}$ \\
      \midrule
      \glsentryshort{nf} & $\begin{cases} 18/25 + \lfloor 100 \cdot (n-1) \rfloor & \text{ if } (n-1) \mod{50} = 0 \\ 5054/75 + \lfloor 100 \cdot (n-1) \rfloor & \text{ if } (n-1) \mod{50} = 33 \\ 9/25 + 2 \cdot (n-1) & \text{ otherwise } \end{cases}$ & $12/5 + 2 \cdot (n-1)$ & $\begin{cases} 42/25 & = 1.68~ms\\76/75 & \simeq 1.01~ms\\51/25 & \simeq 2.04~ms \end{cases}$ \\
      \midrule
      \glsentryshort{imuCorr} & $3/25 + 2 \cdot (n-1)$ & $2 + 2 \cdot (n-1)$ & $47/25 = 1.88~ms$ \\
      \midrule
      \glsentryshort{ncImuInt} & $6/25 + 2 \cdot (n-1)$ & $11/5 + 2 \cdot (n-1)$ & $49/25 = 1.96~ms$ \\
      \midrule
      \glsentryshort{fcImuInt} & $6/25 + 2 \cdot (n-1)$ & $12/5 + 2 \cdot (n-1)$ & $54/25 = 2.16~ms$ \\
      \midrule
      \glsentryshort{imu} & $2 \cdot (n-1)$ & $9/5 + 2 \cdot (n-1)$ & $9/5 = 1.8~ms$ \\
      \midrule
      \glsentryshort{alt} & $20 \cdot (n-1)$ & $11/5 + 20 \cdot (n-1)$ & $11/5 = 2.2~ms$ \\
      \midrule
      \glsentryshort{sp} & $\begin{cases} 21/25 + \lfloor 100 \cdot (n-1) \rfloor & \text{ if } (n-1) \mod{50} = 0 \\ 5063/75 + \lfloor 100 \cdot (n-1) \rfloor & \text{ if } (n-1) \mod{50} = 33 \\ 62/25 + 2 \cdot (n-1) & \text{ otherwise } \end{cases}$ & $13/5 + 2 \cdot (n-1)$ & $\begin{cases} 44/25 & = 1.76~ms \\ 82/75 & \simeq 1.09~ms \\53/25 & =2.12~ms \end{cases}$ \\
      \midrule
      \glsentryshort{ca} & $\begin{cases} 24/25 + \lfloor 100 \cdot (n-1) \rfloor & \text{ if } (n-1) \mod{25} = 0 \\ 3/5 + 8 \cdot (n-1) & \text{ otherwise } \end{cases}$ & $14/5 + 8 \cdot (n-1)$ & $\begin{cases} 46/25 & = 1.84~ms \\ 11/5 & = 1.2~ms \end{cases}$ \\
      \midrule
      \glsentryshort{cy1} & $\begin{cases} 5072/75 + \lfloor 100 \cdot (n-1) \rfloor & \text{ if } (n-1) \mod{25} = 8 \\ 13/5 + 8 \cdot (n-1) & \text{ otherwise } \end{cases}$ & $24/5 + 8 \cdot (n-1)$ & $\begin{cases} 88/75 & = 1.17~ms \\ 11/5 & = 1.2~ms \end{cases}$ \\
      \midrule
      \glsentryshort{ct} & $\begin{cases} 2524/25 + \lfloor 100 \cdot (n-1) \rfloor & \text{ if } (n-1) \mod{25} = 12 \\ 23/5 + 8 \cdot (n-1) & \text{ otherwise } \end{cases}$ & $34/5 + 8 \cdot (n-1)$ & $\begin{cases} 46/25 & = 1.84~ms \\ 11/5 & = 1.2~ms \end{cases}$ \\
      \midrule
      \glsentryshort{cy2} & $\begin{cases} 12572/75 + \lfloor 100 \cdot (n-1) \rfloor & \text{ if } (n-1) \mod{25} = 20 \\ 33/5 + 8 \cdot (n-1) & \text{ otherwise } \end{cases}$ & $44/5 + 8 \cdot (n-1)$ & $\begin{cases} 88/75 & \simeq 1.17~ms \\ 11/5 & = 1.2~ms \end{cases}$ \\
      \midrule
      \glsentryshort{way} & $2 \cdot (n-1)$ & $2 + 2 \cdot (n-1)$ & $2~ms$ \\
      \midrule
      \glsentryshort{mot} & $\begin{cases} 27/25 + \lfloor 100 \cdot (n-1) \rfloor & \text{ if } (n-1) \mod{50} = 0 \\ 5081/75 + \lfloor 100 \cdot (n-1) \rfloor & \text{ if } (n-1) \mod{50} = 33 \\ 1 + 2 \cdot (n-1) & \text{ otherwise } \end{cases}$ & $1 + 2 \cdot (n-1)$ & $\begin{cases} 48/25 & = 1.92~ms \\ 94/75 & \simeq 1.25~ms \\ 2~ms \end{cases}$ \\
      \bottomrule
    \end{tabular}
  }
\end{table*}

\subsection{Feasibility Test of the RMDF Specification of Ingenuity}

\label{sec:ingenuity-feasibility-test}
We have applied the feasibility test presented in~\cite{roumage_realtime_2025} to the Ingenuity specification. This feasibility test asserts that it is necessary that the \gls{wcet} of all actors are less than or equal to the minimum execution windows length presented in column \emph{Exec. Windows} of \cref{tab:ingenuity-timing-constraints} for Ingenuity to be feasible. Those maximum \glspl{wcet} of Ingenuity are presented in \cref{tab:ingenuity-wcet-feasibility}.

\section{A Timing Anomaly on the Sixth Flight of Ingenuity}

\label{sec:timing-anomaly}

During the sixth flight of Ingenuity, an anomaly occurred resulting in a camera frame to be lost\footnote{https://science.nasa.gov/blogs/surviving-an-in-flight-anomaly-what-happened-
on-ingenuitys-sixth-flight/} and Ingenuity tilting back and forth in an oscillating pattern. To the best of our knowledge, this \enquote{lost of camera frame} anomaly is the only public information available about this incident. With the help of the RMDF formalism, this section presents a plausible scenario and the causes that could have led to this incident.

\subsection{Additional Assumptions of Ingenuity's Inner Functioning}

We need to make additional assumptions to specify this anomaly in \cref{fig:ingenuity-specification}:

\begin{enumerate}
  \item We know from the textual specification~\cite{bayard_visionbased_2019} that the vision system executes on a single core, so we may assume that the vision systems run on a preemptive kernel. \label{item:item-1}
  \item We may assume that the \emph{Controlled Joiner} of \cref{fig:ingenuity-specification} was, in fact, a \emph{Joiner} with a naive behavior; that is, it waits on all input channels and sends on its output channel the first token receive. \label{item:item-2}
  \item We may assume that \emph{Feature Match} expects the frame to arrive in the correct order, i.e., the $n$-th frame produced by the camera is the $n$-th frame received by the \emph{Feature Match}. Otherwise, it is considered as an anomaly, and the frame is discarded. \label{item:item-3}
\end{enumerate}

\begin{table}[htbp]
  \caption{Maximum WCETs of the actors of the Ingenuity system that constitute a necessary condition for its feasibility.}
  \label{tab:ingenuity-wcet-feasibility}
  \begin{tabular}{|c|c||c|c|}
    \toprule
    \textbf{Actors} & \textbf{WCET} & \textbf{Actors} & \textbf{WCET} \\
    \midrule
    Camera & 0.53 ms & NC - IMU - Integration & 1.96 ms \\
    Feature Detection & 0.61 ms & FC - IMU - Integration & 2.16 ms \\
    Label Decider & 0.70 ms & IMU & 1.8 ms \\
    Controlled Splitter & 0.57 ms & Altimeter & 2.2 ms  \\
    Feature Tracking & 0.78 ms & State Propagation & 1.09 ms \\
    Filtering Procedure & 0.85 ms & Control Altitude & 1.2 ms \\
    Pseudo Landmarks & 0.97 ms & Control Yaw 1 & 1.17 ms \\
    Controlled Joiner & 0.73 ms & Control Translation & 1.2 ms \\
    Feature Match & 0.93 ms & Control Yaw 2 & 1.17 ms \\
    Navigation Filter & 1.01 ms & Waypoints & 2 ms \\
    IMU-Correction & 1.88 ms & Motors & 1.25 ms \\
    \bottomrule
  \end{tabular}
\end{table}

\subsection{A Plausible Explicability}

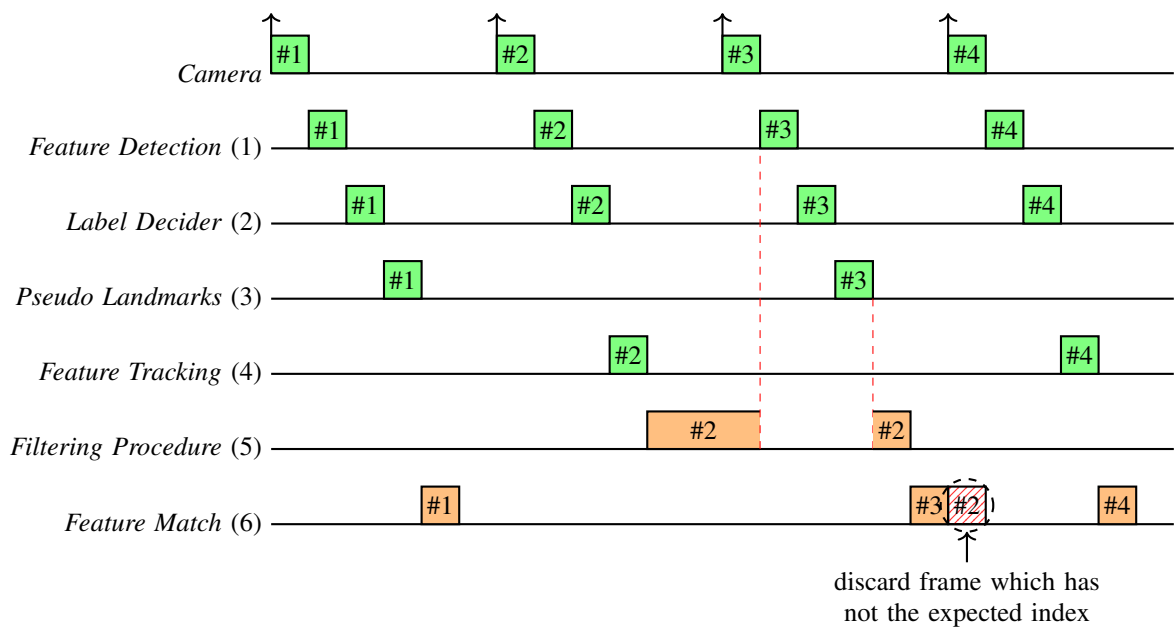
\begin{figure*}[htbp]
  \centering
  \begin{tikzpicture}
    \foreach \x in {0,1,...,6}
      \draw[black, thick] (0,\x) -- (12,\x);
    \foreach \x in {0,3,...,10}
      \draw[black,thick,->] (\x,6) -- (\x,6.8);
    \foreach \x in {1,2,3,4}
      \draw[black,thick,fill=green!50] (\x*3-3,6.5) rectangle (\x*3-2.5,6) node[pos=.5] {\#\x};
    \foreach \x in {1,2,3,4}
      \draw[black,thick,fill=green!50] (\x*3-2.5,5.5) rectangle (\x*3-2,5) node[pos=.5] {\#\x};
    \foreach \x in {1,2,3,4}
      \draw[black,thick,fill=green!50] (\x*3-2,4.5) rectangle (\x*3-1.5,4) node[pos=.5] {\#\x};
    \foreach \x in {1,3}
      \draw[black,thick,fill=green!50] (\x*3-1.5,3.5) rectangle (\x*3-1,3) node[pos=.5] {\#\x};
    \foreach \x in {2,4}
      \draw[black,thick,fill=green!50] (\x*3-1.5,2.5) rectangle (\x*3-1,2) node[pos=.5] {\#\x};
    \node[left] at (0,6) {\emph{Camera}};
    \node[left] at (0,5) {\emph{Feature Detection} (1)};
    \node[left] at (0,4) {\emph{Label Decider} (2)};
    \node[left] at (0,3) {\emph{Pseudo Landmarks} (3)};
    \node[left] at (0,2) {\emph{Feature Tracking} (4)};
    \node[left] at (0,1) {\emph{Filtering Procedure} (5)};
    \node[left] at (0,0) {\emph{Feature Match} (6)};
    \draw[black, thick, fill=orange!50] (6.5,1.5) -- (5,1.5) -- (5,1) -- (6.5,1);
    \draw[dashed, red] (6.5,1) -- (6.5,5);
    \draw[dashed, red] (8,1) -- (8,3);
    \draw[black, thick, fill=orange!50] (8,1) -- (8.5,1) -- (8.5,1.5) -- (8,1.5);
    \draw[black,thick,fill=orange!50] (2,0.5) rectangle (2.5,0) node[pos=.5] {\#1};
    \draw[black,thick,fill=orange!50] (8.5,0.5) rectangle (9,0) node[pos=.5] {\#3};
    \draw[black,thick,pattern=north east lines,pattern color=red!80] (9,0.5) rectangle (9.5,0) node[pos=.5] {\#2};
    \draw[black,thick,fill=orange!50] (11,0.5) rectangle (11.5,0) node[pos=.5] {\#4};
    \draw[dashed,thick] (9.25,0.25) circle (0.35);
    \node[align=center] (text) at (9.25,-1) {discard frame which has\\not the expected index};
    \draw[->,thick] (text) -- (9.25,-0.10);
    \node at (8.25,1.25) {\#2};
    \node at (5.75,1.25) {\#2};
  \end{tikzpicture}
  \caption{Schedule of the actors of the vision processing leading to the anomaly of its sixth flight.}
  \label{fig:ingenuity-sixth-flight-anomaly}
\end{figure*}

The anomaly of the sixth flight is caused by a frame loss. This frame lost could result from a frame inversion followed by a frame discard. Let us assume that during the sixth flight, a job of the actor \emph{Filtering Procedure} takes a longer execution time than its static WCET. The schedule of \cref{fig:ingenuity-sixth-flight-anomaly} presents this situation. A longer execution time of the second job of \emph{Filtering Procedure} implies its preemption (assumption from \cref{item:item-1}), and it implies a frame inversion between frames number 2 and 3, leading to the discard of frame 2. This inversion is because the Joiner waits on all its input channels and sends the first token received (assumption of \cref{item:item-2}). According to \cref{fig:ingenuity-sixth-flight-anomaly}, \emph{Pseudo Landmarks} produces two frames before \emph{Filtering Procedure} produces one. As frame number 2 arrives after frame number 3, frame number 2 is discarded (assumption from \cref{item:item-3}). According to our assumptions, this frame inversion followed by a frame discard may be the anomaly that occurred during the sixth flight of Ingenuity.

\section{Conclusion and Future Works}

\label{sec:conclusion}

We have presented the translation from the textual specification of the Ingenuity Mars helicopter to an RMDF specification. RMDF is a formalism presented in~\cite{roumage_realtime_2025} which is well-suited to specify and analyze mode-dependent cyber-physical systems under relaxed real-time constraints such as Ingenuity. A discussion on the fidelity of the Ingenuity specification was presented. We have also presented a plausible scenario and the causes that could have led to the anomaly of the sixth flight of Ingenuity. As future work, we want to focus on the runtime verification of Ingenuity. We will use the results \cref{tab:ingenuity-timing-constraints} to monitor the timing behavior of Ingenuity during its execution.

\bibliographystyle{IEEEtran}
% \bibliography{IEEEabrv, ../resources/bibliography.bib}

\end{document}